\documentclass[journal=jacsat,manuscript=article]{achemso}

\usepackage[version=3]{mhchem} 
\usepackage{float}

\author{Laura N. Casses}
\affiliation[DTU]
{Department of Electrical and Photonics Engineering, Technical University of Denmark, 2800 Kongens Lyngby, Denmark}
\alsoaffiliation[Nanophoton]
{Centre for Nanophotonics, Technical University of Denmark, 2800 Kongens Lyngby, Denmark}
\alsoaffiliation[CNG]
{Centre for Nanostructured Graphene, Technical University of Denmark, 2800 Kongens Lyngby, Denmark}

\author{Binbin Zhou}

\author{Qiaoling Lin}
\alsoaffiliation[Nanophoton]
{Centre for Nanophotonics, Technical University of Denmark, 2800 Kongens Lyngby, Denmark}
\alsoaffiliation[CNG]
{Centre for Nanostructured Graphene, Technical University of Denmark, 2800 Kongens Lyngby, Denmark}

\author{Annie Tan}

\author{Diane-Pernille Bendixen-Fernex de Mongex}
\author{Korbinian J. Kaltenecker}
\altaffiliation{Current address: Attocube Systems AG, Eglfinger Weg 2, 85540 Haar, Germany}

\author{Sanshui Xiao}
\alsoaffiliation[Nanophoton]
{Centre for Nanophotonics, Technical University of Denmark, 2800 Kongens Lyngby, Denmark}
\alsoaffiliation[CNG]
{Centre for Nanostructured Graphene, Technical University of Denmark, 2800 Kongens Lyngby, Denmark}

\author{Martijn Wubs}
\alsoaffiliation[Nanophoton]
{Centre for Nanophotonics, Technical University of Denmark, 2800 Kongens Lyngby, Denmark}
\alsoaffiliation[CNG]
{Centre for Nanostructured Graphene, Technical University of Denmark, 2800 Kongens Lyngby, Denmark}

\author{Nicolas Stenger}
\email{niste@dtu.dk}

\affiliation[DTU]
{Department of Electrical and Photonics Engineering, Technical University of Denmark, 2800 Kongens Lyngby, Denmark}
\alsoaffiliation[Nanophoton]
{Centre for Nanophotonics, Technical University of Denmark, 2800 Kongens Lyngby, Denmark}
\alsoaffiliation[CNG]
{Centre for Nanostructured Graphene, Technical University of Denmark, 2800 Kongens Lyngby, Denmark}

\title[An \textsf{achemso} demo]
  {Full quantitative near-field characterization of strongly coupled exciton-plasmon polaritons in thin-layered WSe$_2$ on a monocrystalline gold platelet}


\begin{document}


\begin{abstract}
Exciton-plasmon polaritons (EPPs) are attractive both for the exploration of fundamental phenomena and applications in nanophotonics.
Previous studies of EPPs mainly relied on far-field characterization. 
Here, using near-field optical microscopy, we quantitatively characterize the dispersion of EPPs existing in 13-nm-thick  tungsten diselenide (WSe$_2$) deposited on a monocrystalline gold platelet.
We extract from our experimental data a Rabi splitting of 81 meV, and an experimental effective polariton loss of 55 meV, demonstrating that our system is in the strong-coupling regime. 
Furthermore, we measure for the first time at visible wavelengths the propagation length of these EPPs for each excitation energy of the dispersion relation. 
To demonstrate the quantitative nature of our near-field method to obtain the full complex-valued wavevector of EPPs, we use our near-field measurements to predict, via the transfer matrix method, the far-field reflectivities across the exciton resonance. These predictions are in excellent agreement with our experimental far-field measurements. 
Our findings open the door towards the full near-field study of light-manipulating devices at the nanoscale.
\end{abstract}

\section{Introduction}
Near-field microscopy is a powerful and versatile tool that has been used for the direct observation of polaritons in different structures across a wide range of excitation energies at the nanoscale \cite{Chen2012optical, Andryieuski2014direct,Hu2017imaging,Hu2019imaging,Menabde2022near}, as well as the study of the local dielectric function of various materials \cite{Govyadinov2013quantitative,Tranca2015high,Ruta2020quantitative,Alvarez2020infrared,Zhang2022nano}. Many studies using a scattering-type scanning near-field optical microscope (s-SNOM) \cite{Keilmann2004near} have focused on the quantitative near-field characterization of polaritons in the mid-infrared wavelength range\cite{Fei2012gate,Dai2014tunable,Yoxall2015direct,Ma2018plane,Taboada2020broad,Bylinkin2021real}. In contrast, the full s-SNOM characterization of polaritons in the visible to near-infrared range remains unexplored. 
The confinement of polaritons in this range is indeed moderate compared to other polaritons, such as plasmons in graphene \cite{Woessner2015highly}. Thus, in a reflection configuration, the polaritons can be launched both by the s-SNOM tip and by the sharp edge of metallic samples. As shown in previous works \cite{Walla2018anisotropic,Kaltenecker2020mono,Casses2022quantitative}, this leads to three main excitation channels for the polaritons, namely the tip-launched, the edge-launched, and the tip-reflection edge-launched polaritons. The plurality of possible excitation paths leads to complex interference patterns, which complicates the analysis of the measurements. 

As many excitonic excitations have energies in the visible and near-infrared range~\cite{Basov2016polaritons,Hu2017imaging,Low2017polaritons,Basov2021polariton}, the study and full quantitative characterization of polaritons in this range are highly relevant for the future development of the field. 
In particular, exciton-plasmon polaritons (EPPs) are hybrid states emerging from the coupling between excitons \cite{Mueller2018exciton} and surface plasmon polaritons (SPPs) \cite{Maier2007plasmonics}. These polaritons have the potential to become a building block for applications \cite{Xia2014two,Baranov2018novel} such as nanoscale polaritonic lasers \cite{Byrnes2014exciton,Sanvitto2016road}, and new quantum emitters for quantum information processing \cite{Denning2020cavity}. A pioneering work measured EPPs resulting from the interaction between excitons in  tungsten diselenide (WSe$_2$) and SPPs on gold, for thicknesses of WSe$_2$ ranging from 35 to 110 nm \cite{Iyer2022nano}. While showing impressive confinement of the EPPs at higher thicknesses, the high absorption losses associated with this thick excitonic material hindered the retrieval of the upper polariton branch. Thus, the Rabi splitting energy was obtained from numerical calculations. In addition, the propagation length of the EPPs, directly related to the amount of radiative and non-radiative losses present in the polaritons \cite{Jang2024fourier}, was not determined. This is however a crucial information to be retrieved experimentally in order to assess the performances of a polaritonic device. 

In our previous works \cite{Kaltenecker2020mono,Casses2022quantitative}, we showed that by carefully choosing the azimuthal angle of incidence relative to the edge of monocrystalline gold platelets, we could isolate the excitation path of interest, in this case the tip-launched SPP excitation path. Using Fourier analysis, we could retrieve both the wavelength and the propagation length of moderately confined SPPs on monocrystalline gold platelets. 

In this work, we study EPPs propagating in a 13-nm flake of WSe$_2$ deposited on a monocrystalline gold platelet. We use a s-SNOM in the reflection configuration~\cite{Casses2022quantitative} to measure the edge-launched EPP excitation path for excitation energies around the A-exciton energy of WSe$_2$ (1.63 eV). The choice of perpendicular incidence and a defocused beam allows us to measure a clear signal coming from the EPPs. With this information, we are able to retrieve with good accuracy the experimental dispersion relation, for the lower as well as for the upper polariton branches. As we can resolve both branches, we are able to determine a Rabi splitting based on experimental values only, without using numerical calculations or simulations. Furthermore, we determine the propagation lengths of the EPPs for all the measured energies. Thereby, we reconstruct the full complex-valued polariton wavevector, giving us access to the fundamental properties of EPPs, such as the effective losses of the hybridized light-matter excitation. The experimental determination of the polariton losses shows us that our system is in the strong-coupling regime. In addition, the near-field experimental values allow us to determine the in-plane dielectric function of WSe$_2$. These values, obtained from the near-field measurements, are subsequently used to predict the far-field reflectivity of our sample. We find good agreement with the reflectivity measured directly with a far-field microscope. 

This work reports the first quantitative full near-field characterization of the nanoscale properties of strongly coupled light-matter states in the visible to near-infrared energy range, the relevant range for most EPPs found in nature. The methodology presented here pushes forward the analytical capabilities of state-of-the-art s-SNOM allowing the full retrieval of the coupling strength, propagating losses and effective losses of EPPs. Furthermore, our results show that it is possible to relate near-field information with far-field optical properties. This link between the near- and far-field will be crucial in the design of new devices to manipulate light and excitons on the nanoscale for future low-threshold nano-lasers, new fast quantum gates, quantum emitters integrated in nanocavities, as well as the development of technology-relevant metasurfaces.

\section{Results and discussion}

\subsection{Near-field measurements of the EPPs}

\begin{figure}[h]
\centering
\includegraphics[width=\textwidth]{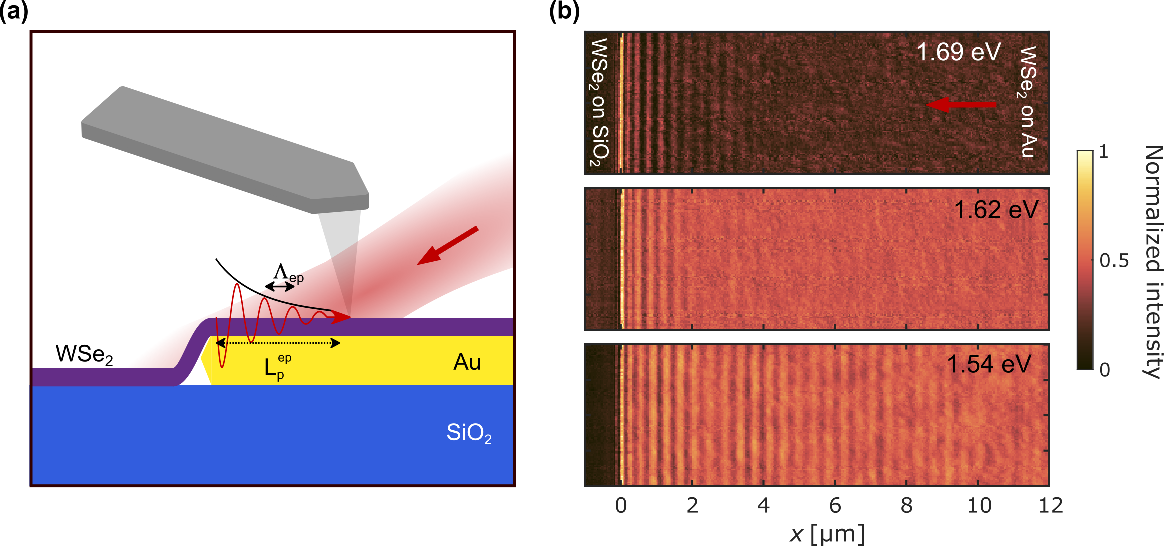}
\caption{Near-field measurement of exciton-plasmon polaritons (EPPs) propagating on a thin layer of WSe$_2$ deposited on a monocrystalline gold platelet. (a)~Illustration of the experiment. Here, $\Lambda_{\text{ep}}$ is the wavelength of the edge-launched EPPs, and $L_\text{p}^{\text{ep}}$ is the propagation length of the edge-launched EPPs. The 2-nm layer of aluminum oxide between the gold and WSe$_2$ is not represented here. (b)~Near-field intensity at three different excitation energies. The origin of the $x$-axis is taken at the edge of the gold platelet, and the red arrow represents the direction of the incident light.}\label{fig1}
\end{figure}

A sketch of the near-field experiment is presented in Figure \ref{fig1}a. To investigate the properties of EPPs in the near-field, we use a 
100-nm thick monocrystalline gold platelet on silicon dioxide, protected by a 2-nm layer of aluminum oxide. These platelets are similar to the ones studied in previous works \cite{Krauss2018,Kaltenecker2020mono,Casses2022quantitative}. A 13-nm thick flake of WSe$_2$ is mechanically exfoliated and transferred \cite{Castellanos2014deterministic} (see Methods) onto the gold platelet such that it covers the edge of the platelet. 
The EPPs are measured using a s-SNOM in the reflection configuration \cite{Kaltenecker2020mono} (see Methods). A tunable laser is focused on the s-SNOM tip such that the laser beam is perpendicular to the edge of the platelet in the plane of the sample (see red arrow in Figure \ref{fig1}b). 
This configuration allows to excite EPPs of wavelength $\Lambda_{\text{ep}}$ and propagation length L$_\text{p}^{\text{ep}}$ at the sharp edge of the monocrystalline platelet, for excitation energies from 1.5 to 1.7 eV.

The experiments are performed such that the sample is scanned under the fixed combination of the oscillating tip and excitation light always focused on the tip apex. With a highly focused beam, the excitation of the edge-launched EPPs would therefore depend on the tip-edge position and thus influence the determination of L$_\text{p}^{\text{ep}}$. 
To overcome this challenge, we slightly defocus the beam in front of the s-SNOM (see Methods), as in Ref. \citenum{Andryieuski2014direct}. The resulting beam diameter at the tip apex is estimated to be about 50~\textmu m. 

Figure \ref{fig1}b shows three examples of the near-field intensity obtained with our measurements, below (1.54 eV), around (1.62 eV), and above (1.69 eV) the A-exciton energy. The maps from all the measurements can be found in the Supplementary Information. The edge of the platelet corresponds to the origin of the $x$-axis. The interference patterns seen on the maps and parallel to the edge correspond to the interference between the edge-launched EPPs and the background light scattered directly from the sample \cite{Walla2018anisotropic,Kaltenecker2020mono}. The wavevector of the interference fringes $K_{\text{el}}$ can be expressed as \cite{Walla2018anisotropic}
\begin{equation} \label{eq1}
    K_{\text{el}}(\theta, \varphi) =  \frac{2\pi}{\lambda_0}\left(-\sin(\theta)\sin(\varphi)+\sqrt{\sin^2(\theta)\sin^2(\varphi)-\sin^2(\theta)+n^2}\right), 
\end{equation}
where $\lambda_0$ is the wavelength of the incident light, $\theta = 60^\circ$ is the polar angle between the incident light and the surface of the sample, $\varphi = -90^\circ$ is the azimuthal angle between the incident light and the edge of the sample, and $n$ is the real part of the effective refractive index of the confined mode of the multilayer structure. At perpendicular incidence, Equation~\ref{eq1} becomes equivalent to the expression for the wavelength of the interference fringes in Ref. \citenum{Hu2017imaging}.

It should be highlighted that $K_{\text{el}}$ is the wavevector of the interference fringes obtained with the near-field measurements. Due to the interference with the background scattered light, $K_{\text{el}}$ is different from the wavevector of the confined EPPs mode, $K_{\text{ep}}~=~2\pi n / \lambda_0$. The wavevector of the EPPs $K_{\text{ep}}$ can be calculated as 
\begin{equation} \label{eq2}
    K_{\text{ep}}=  K_{\text{el}} \frac{n}{-\sin(\theta)\sin(\varphi)+\sqrt{\sin^2(\theta)\sin^2(\varphi)-\sin^2(\theta)+n^2}}. 
\end{equation}

The energy dependence of the propagation length is already apparent from the near-field intensity maps, with a shorter propagation length above than below the exciton energy, due to the additional absorption losses of WSe$_2$. As the incident light has an azimutal angle $\varphi$ perpendicular to the edge, the propagation length of the interference fringes is equivalent to the propagation length of the EPPs: $L_\text{p}^\text{el} = L_\text{p}^\text{ep}$. 

\subsection{Real-space and Fourier analysis of the EPPs}

\begin{figure*}[h]
\centering
\includegraphics[width=\textwidth]{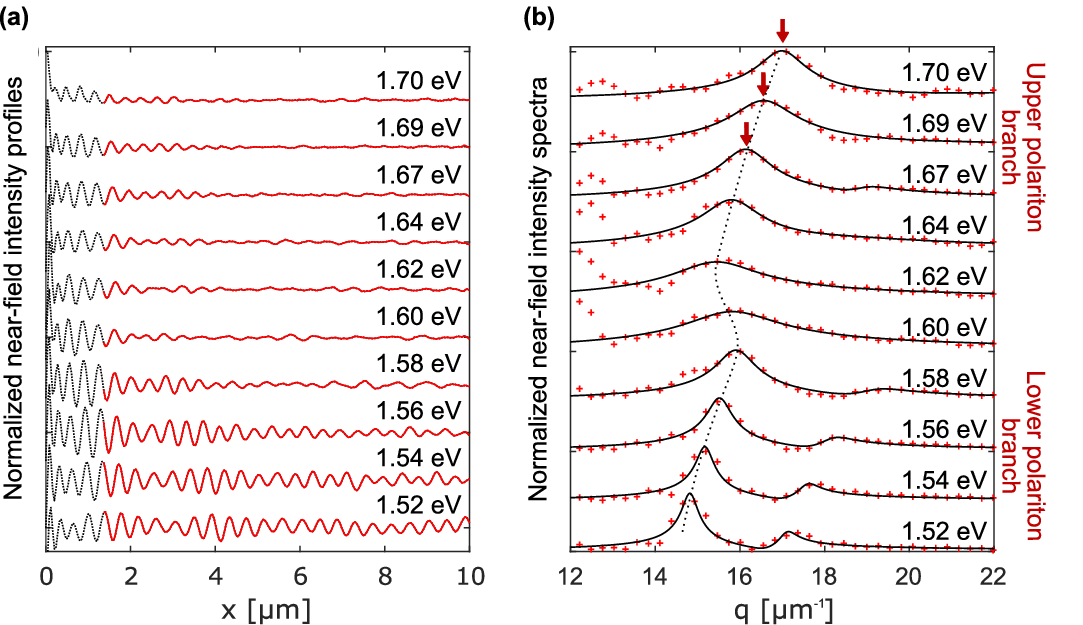}
\caption{(a)~Fringe profiles on the WSe$_2$/Al$_2$O$_3$/gold structure at the different excitation energies measured with the s-SNOM. The red sections of the profiles are used for the Fourier transform. (b)~Fourier transforms of the profiles (red crosses), and their fits (black curves). The red arrows highlight the position of the edge-launched polaritons peaks of wavevector $K_\text{el}$. The black dotted line is a guide to the eye.}\label{fig2}
\end{figure*}

Figure \ref{fig2}a presents the profiles made by averaging the experimental maps in the direction parallel to the edge, for all the measured excitation energies. Above the A-exciton energy - corresponding to the upper polariton branch - exponentially decaying interference patterns can be seen up to a distance of about 2 to 4 \textmu m, depending on the excitation energy. 
Below the exciton energy, the propagation lengths are longer, and the interference patterns are more complex. The fringes indeed display a revival behavior, hinting to the contribution from two different excitation paths. 

For a better analysis of the wavevector components at different energies, we apply the Fourier transform on the profiles, using the same procedure as our previous work~\cite{Casses2022quantitative}. The first 1.4~\textmu m of the profiles (see the black dotted lines in Figure \ref{fig2}a) induce a background interfering with the rest of the signal, possibly due to multiple scattering between the tip and the sample at the edge of the platelet. Thus, the sections from $x=0$~\textmu m to $x=1.4$~\textmu m are removed for the Fourier transform. 
The resulting spectra are shown in Figure \ref{fig2}b. The peak corresponding to the first type of edge-launched EPPs \cite{Kaltenecker2020mono} is indicated by the red arrows and the dotted line. 
As the energy varies, the peak positions present a sizeable back-bending indicating coupling between the SPPs and the A-exciton. Above the exciton energy, mainly one peak can be seen in each spectrum. A second peak can sometimes be seen around 12 \textmu m$^{-1}$, and has a wavevector corresponding to the tip-reflected edge-launched polaritons excitation path \cite{Walla2018anisotropic, Kaltenecker2020mono}. 
Below the exciton energy, a small peak around 17-19 \textmu m$^{-1}$ can also be observed. This peak corresponds to the tip-launched EPPs excitation path. 
The revival of the interference fringes at lower energies is thus due to the interference between tip-launched and edge-launched EPPs. 
The peak due to the tip-launched EPPs cannot be seen at higher energies, because the tip-launched EPPs travel twice the distance compared to the edge-launched EPPs. Hence, their signal becomes too weak due to higher losses in the material. 

To extract the value of the EPP wavevector as a function of the energy, we fit the spectra as the sum of two Lorentzian functions. We use Lorentzian functions since edge-launched polaritons can be well approximated by plane waves \cite{Kaltenecker2020mono,Casses2022quantitative}. The results of the fitting are presented as the black lines in Figure \ref{fig2}b. 
The fit of the spectra allows to retrieve the wavevector $K_\text{el}$ and the propagation length $L_\text{p}^\text{el}$ of the edge-launched EPP interference patterns. However, the interference profiles have a finite interval length, which implies for the Fourier transform that our signal is inherently multiplied with a square window. Thus, each spectral shape is the convolution between a Lorentzian function and a sinc function \cite{Bracewell1986Fourier}. Therefore, while the Fourier transform gives an accurate value for the wavevectors, the accurate determination of the propagation length is hindered. 
To fit the propagation length with a higher accuracy, we directly fit the interference profiles from Figure \ref{fig2}a. To help the fitting procedure, the value of the wavevectors are fixed and taken from the fitting results of the Fourier transformed spectra. Our analysis thus makes use of the advantages of both the Fourier and the real spaces.

\subsection{Reconstruction of the full complex-valued wavevector}

\begin{figure*}[h!]
\centering
\includegraphics[width=0.83\textwidth]{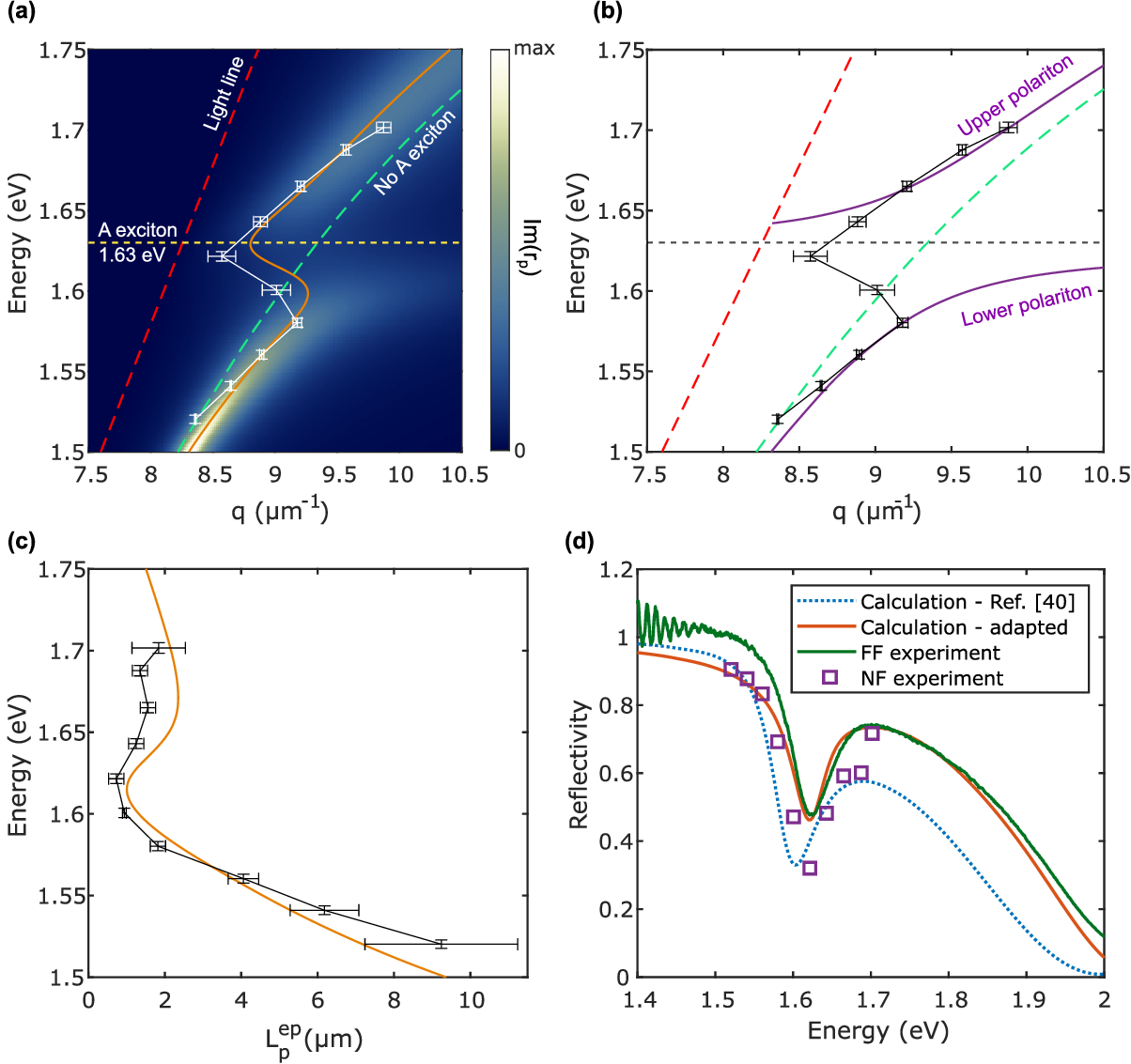}
\caption{Retrieval of the EPP properties. (a) Dispersion relation of the exciton-plasmon polaritons for a 13-nm-thick WSe$_2$ deposited on gold. The colormap displays the imaginary part of the reflection coefficient, calculated with the TMM. The white line with error bars corresponds to the experimental wavevector extracted from the data. The vertical error bars correspond to the linewidth of the incident laser, and the horizontal ones are the fitting uncertainty on the peak position. The orange line to the theoretical dispersion relation calculated using the TMM.  The red dashed line represents the light line in air, the horizontal dashed line the energy of the A-exciton of WSe$_2$, and the blue dashed line the dispersion of the sample without the A-exciton. (b)~Dispersion relation of the EPPs compared to the coupled-oscillator model (COM). The two polariton branches are plotted in purple, and the experimental wavevector is in black. (c)~Experimental (black curve) and theoretical (orange curve) propagation lengths of the polaritons. The horizontal error bars correspond to the uncertainty of the fit. (d)~Comparison between the reflectivity of 13-nm-thick WSe$_2$ on gold, calculated using the WSe$_2$ dielectric function from Munkhbat et al. \cite{Munkhbat2022optical} (blue dotted line), measured directly with a conventional far-field microscope (green line), calculated using a dielectric function adapted to fit the far-field reflectivity (red line), and extracted from the near-field measurements (purple squares).}\label{fig3}
\end{figure*}

The wavevectors of the EPPs are calculated from the retrieved values of $K_\text{el}$ using Equation~\ref{eq2}. The experimental dispersion relation resulting from this process is depicted as the white line in Figure \ref{fig3}a. The vertical error bars correspond to the linewidth of the incident laser, and the horizontal ones are the fitting uncertainty on the peak position. To compare our experimental results with theory, we use the transfer matrix method (TMM) adapted to the case of an anisotropic material. Since WSe$_2$ is a uniaxial anisotropic crystal, we use the $2\times2$ matrices presented in Ref. \citenum{Majerus2018electrodynamics} in the case of \textit{p}-polarization. For our calculations, we use values of the dielectric function of gold from McPeak \textit{et al.} \cite{McPeak2015}, as it has been shown to have good agreement with the dielectric function of monocrystalline gold \cite{Casses2022quantitative,Lebsir2022ultimate}. The dielectric function of aluminum oxide is set at 2.56. To obtain the dielectric function of WSe$_2$, we first fit the WSe$_2$ dielectric function data from the tabulated values of Munkhbat \textit{et al.} \cite{Munkhbat2022optical}. 
Then, we measure with a far-field microscope (see Methods) the reflectivity of our multilayer structure, and compare it with the reflectivity calculated using the values from Ref. \citenum{Munkhbat2022optical}. The experimental and calculated reflectivities differ, as can be seen in Figure \ref{fig3}d. This difference in reflectivities hints at a difference in the in-plane dielectric function of WSe$_2$. There could be several reasons for this difference, as the measurement of the dielectric function of WSe$_2$ in Ref. \citenum{Munkhbat2022optical} was performed on a thicker WSe$_2$ flake deposited on silicon dioxide. To assess the influence of the substrate, we also measure the reflectivity of our WSe$_2$ flake on the silicon dioxide substrate next to the gold platelet. The comparison between the reflectivities of WSe$_2$ on the monocrystalline gold platelet and on silicon dioxide can be found in the Supplementary Information. This comparison shows that the reflectivity of our WSe$_2$ flake on silicon dioxide follows much better the reflectivity calculated using the values from Ref. \citenum{Munkhbat2022optical}. Thus, we attribute the difference in reflectivity of WSe$_2$ mainly to a change in the in-plane dielectric constant of WSe$_2$ due to the presence of a metallic substrate. To take this change into account, we adapt the parameters from the in-plane WSe$_2$ dielectric function fitted based on Ref. \citenum{Munkhbat2022optical} to correspond better to our experimental reflectivity. The dielectric function of WSe$_2$ so obtained can be found in the Supplementary Information. 

The theoretical dispersion relation, plotted as the orange line in Figure \ref{fig3}a, is calculated by finding the poles of the reflection coefficient $r_p$ of our multilayer structure \cite{Zhan2013transfer}. The numerical solution to this first approach is a complex-valued wavevector, from which the real part gives the dispersion relation. The colour plot shows, as a second approach, the imaginary part of the reflection coefficient Im($r_p$) - which has been linked to the optical density of states (DOS)\cite{Hu2017imaging,Hu2019imaging,Iyer2022nano}. The two approaches both give consistent results and show a clear agreement with the experimental results.

To characterize the Rabi splitting, we first consider the difference between the upper and lower polariton branches at zero detuning, with zero detuning defined as the intersection of the exciton mode and the surface plasmon mode without coupling to the A-exciton of WSe$_2$ \cite{Geisler2019single}. The dispersion of these two modes is shown as the yellow and blue dashed lines in Figure \ref{fig3}a, respectively. The vertical cross-section of the theoretical DOS at the zero detuning ($q_0 = 9.4$~\textmu m$^{-1}$) gives a value for the Rabi splitting energy of $E_{\text{Rabi}_{\text{0}}} = 89$ meV. This method has, however, the disadvantage of being correlated with the experiment in an indirect manner, as the Rabi splitting energy is measured from the theoretical calculations and the comparison between the theory and the experimental data is here qualitative. 
In our case, the center of the experimental and theoretical back-bendings are considerably shifted compared to the zero detuning determined by the method explained above. This effect has also been observed in a previous theoretical study \cite{Goncalves2018plasmon}. Adding or removing the A-exciton resonance for a 13-nm thick WSe$_2$ has indeed a stronger influence on the WSe$_2$ dielectric function overall, and thus on the confinement of the surface plasmons, compared to the monolayer case. Thus, the determination of the zero detuning condition is difficult since it depends strongly on the thickness of the excitonic material used and on the retrieval of the A-exciton. A method relying only on experimental data is therefore preferable to find the amplitude of the Rabi splitting. 
At the center of the theoretical back-bending ($q_1 = 9.0$~\textmu m$^{-1}$), we find a value for the Rabi splitting of $E_{\text{Rabi}_{\text{1}}} = 77$ meV, while $E_{\text{Rabi}_{\text{2}}} = 83$ meV at the center of the experimental back-bending ($q_2 = 8.9$~\textmu m$^{-1}$). By comparison, fitting our experimental dispersion data with the coupled-oscillator model (COM) \cite{Goncalves2018plasmon} (see Figure \ref{fig3}b) gives a value of $E_{\text{Rabi}_{\text{4}}} = 81$ meV. In the following we will use the more conservative value of the Rabi splitting determined experimentally, i.e. 81 meV. The experimentally and theoretically determined values of the Rabi splitting fall all within a 10 percent interval around this value, showing the excellent agreement between experiment and theory. 


As the solutions to finding the poles of the reflection coefficients are complex-valued wavevectors, the theoretical propagation length can also be deduced from the imaginary part of these wavevectors. The theoretical and experimental propagation lengths are plotted as a function of the energy in Figure \ref{fig3}c. The experimental propagation length is found by fitting the profiles in real space and clearly follows the theoretical calculations.  

With the characterized dispersion relation and the propagation lengths, we can reconstruct the energy dependence of the complex-valued wavevector from near-field measurements of EPPs in the optical energy range, for the first time to our knowledge. This allows us to calculate the effective dielectric function of the EPPs, relying only on experimental measurements. The EPP effective dielectric function is plotted in the Supplementary Information, as a function of the excitation energy. The imaginary part of this dielectric function is associated with the polariton losses. By fitting the polariton dielectric function as the sum of two Lorentz oscillators, the polariton losses are found to be equal to $\Gamma=55$~meV. Thus, we can infer that our system is in the strong-coupling regime, relying only on near-field experimental results, without inference via theoretical calculations or simulations. 

As a benchmark of the quality of our data, we compare the reflectivity measured with a far-field microscope with the reflectivity predicted from the near-field measurements. We use the TMM to find the values of the in-plane WSe$_2$ dielectric function from the experimental values of the complex-valued wavevector $\Tilde{K}_\text{ep}=K_\text{ep} + i/L_\text{p}^\text{ep}$. The new values of the WSe$_2$ dielectric function are then used to calculate the reflectivity. The reflectivity recalculated from the near-field data shows a clear dip due to the A-exciton absorption, and agrees well with the direct far-field measurement of the reflectivity (see Figure \ref{fig3}d). Interestingly, almost all the data points from the near-field experiments lie in between the reflectivity calculated using directly the data from Ref.~\citenum{Munkhbat2022optical} and the reflectivity calculated using the adapted in-plane WSe$_2$ dielectric function (see Supplementary Information). Our procedure highlights the correspondence between material parameters obtained from polaritonic near-field measurements and the ones obtained with far-field methods. Compared to other s-SNOM studies of the dielectric function \cite{Govyadinov2013quantitative,Tranca2015high,Ruta2020quantitative,Alvarez2020infrared,Zhang2022nano} our method does not extract the dielectric function at the nanoscale, but gives the averaged dielectric function felt by the EPPs. Our method allows us to retrieve the complex-valued dielectric function of the excitonic material in the strong coupling regime and makes the reconstruction of far-field optical properties of nanometer-thin materials possible. 

\section{Conclusion}

In this work, we used a s-SNOM in the reflection configuration to measure EPPs on a multilayer structure made of gold and 13-nm thick WSe$_2$, at different energies around the A-exciton energy.
By expanding the incident laser beam and carefully choosing the direction of the incident light, we could resolve both the higher and lower polariton branches of the edge-launched EPP excitation path.
By using a combination of Fourier and real-space analysis, we characterized the wavevectors and, for the first time to our knowledge, the propagation lengths of the EPPs. We observed a clear back-bending of the dispersion relation, hinting at strong coupling. Using the experimental data only, we retrieved the Rabi splitting energy of $E_{\text{Rabi}} = 81$~meV and polaritonic losses of $\Gamma=55$~meV, thus demonstrating that our structure is indeed in the strong-coupling regime. Furthermore, our analysis allowed us to highlight the importance of the substrate for the accurate determination of the in-plane 
dielectric constant of thin-layered WSe$_2$. The quality of the data was then demonstrated by predicting the far-field reflectivity of our sample based on the values obtained through the near-field measurements, with excellent agreement with the direct measurement of far-field reflectivity. The methods developed in this work can be applied to the quantitative characterization of the complex-valued wavevectors of different types of polaritons, for a s-SNOM in a reflection configuration.  It can thus be used for the investigation of quantum related phenomena in polariton physics \cite{Basov2021polariton,Boroviks2022extremely}, at the nanometer scale and at visible frequencies. Moreover, relating the far-field and the near-field information is crucial in the design of new polaritonic devices, in the understanding of material properties in the strong-coupling regime, as well as in the design of new metasurfaces.

\section{Experimental methods}
\noindent \textbf{Sample preparation}: A WSe$_2$ crystal (HQ graphene) was mechanically exfoliated with a light blue tape (1007R, SPS Ltd). A 13-nm-thick WSe$_2$ flake resulting from the exfoliation was then transferred onto the monocrystalline gold platelets (NanoStruct GmbH), using the dry-transfer method \cite{Castellanos2014deterministic}.

\noindent \textbf{Near-field measurements}: The measurements were performed with a commercial s-SNOM in reflection mode (neaSNOM, Attocube Systems AG). The polaritons were excited with a \textit{p}-polarized femtosecond Ti:sapphire laser (Tsunami Model 3960, Spectra-Physics Inc.) and the near-field signal was recorded using a 10-MHz adjustable photodiode (2051, Newport Corp.). 
The laser was directed towards a platinum-coated AFM tip (Arrow-NCPt, NanoAndMore GmbH) with a nominal apex radius of 25~nm. The tip is oscillating above the sample at a frequency of $f_0 \sim 300$ kHz and with an oscillation amplitude of about 40 nm. The signal recorded by the photodiode is demodulated at $3 f_0$.

To defocus the beam spot at the tip apex, a combination of a divergent lens of focal length $f_d = -200$ mm and a convergent lens $f_c = 300$ mm is placed - 5 cm apart - in front of the s-SNOM setup. With this effective divergent lens, the diameter of the spot at the tip is calculated to be 53 \textmu m. 

\noindent \textbf{Far-field measurements}: The reflection measurements were performed with a customised microscopy setup based on an inverted microscope
(Eclipse Ti-U, Nikon). A white light source was focused onto the sample with a 10X objective. The collected light was extracted with the same objective and sent to a spectrometer (Andor Solis 303I).

\begin{acknowledgement}
LNC thanks K. R. Spiegelhauer for his help with preliminary s-SNOM experiments, B. Munkhbat for providing the data of the WSe$_2$ dielectric constant now published in Ref.~\citenum{Munkhbat2022optical}, and V. Zenin, P. Bøggild and S. Raza for valuable discussions.  LNC, SX, MW, and NS acknowledge support from Danmarks Grundforskningsfond (Grants No. DNRF147 - NanoPhoton and No. DNRF103 - CNG); Parts of our optical setup are financed through the IDUN Center of Excellence founded by the Danish National Research Foundation (project number DNRF122) and VILLUM FONDEN (project number 9301); NS acknowledges Villum Fonden (00028233); NS and MW acknowledge support from Det Frie Forskningsråd, Natur og Univers (0135-00403B); QL and SX acknowledge support from Det Frie Forskningsråd, Teknologi og Produktion (9041-00333B); LNC and NS acknowledge support from Det Frie Forskningsråd, Teknologi og Produktion (1032-00496B).



\end{acknowledgement}

\begin{suppinfo}

Supplementary information is available for this article. It includes the near-field intensity maps at all measured energies, the real-space fit of the near-field profiles, and additional details on the analysis of the dielectric function of WSe$_2$ profiles and the study of the EPP dielectric function. 


\end{suppinfo}

\bibliography{BibliographyLNCasses.bib}

\providecommand{\latin}[1]{#1}
\makeatletter
\providecommand{\doi}
  {\begingroup\let\do\@makeother\dospecials
  \catcode`\{=1 \catcode`\}=2 \doi@aux}
\providecommand{\doi@aux}[1]{\endgroup\texttt{#1}}
\makeatother
\providecommand*\mcitethebibliography{\thebibliography}
\csname @ifundefined\endcsname{endmcitethebibliography}  {\let\endmcitethebibliography\endthebibliography}{}
\begin{mcitethebibliography}{45}
\providecommand*\natexlab[1]{#1}
\providecommand*\mciteSetBstSublistMode[1]{}
\providecommand*\mciteSetBstMaxWidthForm[2]{}
\providecommand*\mciteBstWouldAddEndPuncttrue
  {\def\EndOfBibitem{\unskip.}}
\providecommand*\mciteBstWouldAddEndPunctfalse
  {\let\EndOfBibitem\relax}
\providecommand*\mciteSetBstMidEndSepPunct[3]{}
\providecommand*\mciteSetBstSublistLabelBeginEnd[3]{}
\providecommand*\EndOfBibitem{}
\mciteSetBstSublistMode{f}
\mciteSetBstMaxWidthForm{subitem}{(\alph{mcitesubitemcount})}
\mciteSetBstSublistLabelBeginEnd
  {\mcitemaxwidthsubitemform\space}
  {\relax}
  {\relax}

\bibitem[Chen \latin{et~al.}(2012)Chen, Badioli, Alonso-Gonz{\'a}lez, Thongrattanasiri, Huth, Osmond, Spasenovi{\'c}, Centeno, Pesquera, Godignon, Godignon, Elorza, Camara, de~Abajo, Hillenbrand, and Koppens]{Chen2012optical}
Chen,~J. \latin{et~al.}  Optical nano-imaging of gate-tunable graphene plasmons. \emph{Nature} \textbf{2012}, \emph{487}, 77--81\relax
\mciteBstWouldAddEndPuncttrue
\mciteSetBstMidEndSepPunct{\mcitedefaultmidpunct}
{\mcitedefaultendpunct}{\mcitedefaultseppunct}\relax
\EndOfBibitem
\bibitem[Andryieuski \latin{et~al.}(2014)Andryieuski, Zenin, Malureanu, Volkov, Bozhevolnyi, and Lavrinenko]{Andryieuski2014direct}
Andryieuski,~A.; Zenin,~V.~A.; Malureanu,~R.; Volkov,~V.~S.; Bozhevolnyi,~S.~I.; Lavrinenko,~A.~V. Direct characterization of plasmonic slot waveguides and nanocouplers. \emph{Nano Lett.} \textbf{2014}, \emph{14}, 3925--3929\relax
\mciteBstWouldAddEndPuncttrue
\mciteSetBstMidEndSepPunct{\mcitedefaultmidpunct}
{\mcitedefaultendpunct}{\mcitedefaultseppunct}\relax
\EndOfBibitem
\bibitem[Hu \latin{et~al.}(2017)Hu, Luan, Scott, Yan, Mandrus, Xu, and Fei]{Hu2017imaging}
Hu,~F.; Luan,~Y.; Scott,~M.; Yan,~J.; Mandrus,~D.; Xu,~X.; Fei,~Z. {Imaging exciton--polariton transport in MoSe$_2$ waveguides}. \emph{Nature Photonics} \textbf{2017}, \emph{11}, 356--360\relax
\mciteBstWouldAddEndPuncttrue
\mciteSetBstMidEndSepPunct{\mcitedefaultmidpunct}
{\mcitedefaultendpunct}{\mcitedefaultseppunct}\relax
\EndOfBibitem
\bibitem[Hu \latin{et~al.}(2019)Hu, Luan, Speltz, Zhong, Liu, Yan, Mandrus, Xu, and Fei]{Hu2019imaging}
Hu,~F.; Luan,~Y.; Speltz,~J.; Zhong,~D.; Liu,~C.~H.; Yan,~J.; Mandrus,~D.~G.; Xu,~X.; Fei,~Z. {Imaging propagative exciton polaritons in atomically thin ${\mathrm{WSe}}_{2}$ waveguides}. \emph{Phys. Rev. B} \textbf{2019}, \emph{100}, 121301\relax
\mciteBstWouldAddEndPuncttrue
\mciteSetBstMidEndSepPunct{\mcitedefaultmidpunct}
{\mcitedefaultendpunct}{\mcitedefaultseppunct}\relax
\EndOfBibitem
\bibitem[Menabde \latin{et~al.}(2022)Menabde, Boroviks, Ahn, Heiden, Watanabe, Taniguchi, Low, Hwang, Mortensen, and Jang]{Menabde2022near}
Menabde,~S.~G.; Boroviks,~S.; Ahn,~J.; Heiden,~J.~T.; Watanabe,~K.; Taniguchi,~T.; Low,~T.; Hwang,~D.~K.; Mortensen,~N.~A.; Jang,~M.~S. Near-field probing of image phonon-polaritons in hexagonal boron nitride on gold crystals. \emph{Science Advances} \textbf{2022}, \emph{8}, eabn0627\relax
\mciteBstWouldAddEndPuncttrue
\mciteSetBstMidEndSepPunct{\mcitedefaultmidpunct}
{\mcitedefaultendpunct}{\mcitedefaultseppunct}\relax
\EndOfBibitem
\bibitem[Govyadinov \latin{et~al.}(2013)Govyadinov, Amenabar, Huth, Carney, and Hillenbrand]{Govyadinov2013quantitative}
Govyadinov,~A.~A.; Amenabar,~I.; Huth,~F.; Carney,~P.~S.; Hillenbrand,~R. Quantitative Measurement of Local Infrared Absorption and Dielectric Function with Tip-Enhanced Near-Field Microscopy. \emph{The Journal of Physical Chemistry Letters} \textbf{2013}, \emph{4}, 1526--1531\relax
\mciteBstWouldAddEndPuncttrue
\mciteSetBstMidEndSepPunct{\mcitedefaultmidpunct}
{\mcitedefaultendpunct}{\mcitedefaultseppunct}\relax
\EndOfBibitem
\bibitem[Tranca \latin{et~al.}(2015)Tranca, Stanciu, Hristu, Stoichita, Tofail, and Stanciu]{Tranca2015high}
Tranca,~D.; Stanciu,~S.; Hristu,~R.; Stoichita,~C.; Tofail,~S.; Stanciu,~G. High-resolution quantitative determination of dielectric function by using scattering scanning near-field optical microscopy. \emph{Scientific Reports} \textbf{2015}, \emph{5}, 1--9\relax
\mciteBstWouldAddEndPuncttrue
\mciteSetBstMidEndSepPunct{\mcitedefaultmidpunct}
{\mcitedefaultendpunct}{\mcitedefaultseppunct}\relax
\EndOfBibitem
\bibitem[Ruta \latin{et~al.}(2020)Ruta, Sternbach, Dieng, McLeod, and Basov]{Ruta2020quantitative}
Ruta,~F.~L.; Sternbach,~A.~J.; Dieng,~A.~B.; McLeod,~A.~S.; Basov,~D. {Quantitative nanoinfrared spectroscopy of anisotropic van der Waals materials}. \emph{Nano Letters} \textbf{2020}, \emph{20}, 7933--7940\relax
\mciteBstWouldAddEndPuncttrue
\mciteSetBstMidEndSepPunct{\mcitedefaultmidpunct}
{\mcitedefaultendpunct}{\mcitedefaultseppunct}\relax
\EndOfBibitem
\bibitem[{\'A}lvarez-P{\'e}rez \latin{et~al.}(2020){\'A}lvarez-P{\'e}rez, Folland, Errea, Taboada-Guti{\'e}rrez, Duan, Mart{\'\i}n-S{\'a}nchez, Tresguerres-Mata, Matson, Bylinkin, He, Ma, Bao, Ingnacio~Mart{\'\i}n, Caldwell, Nikitin, and Alonso-Gonz{\'a}lez]{Alvarez2020infrared}
{\'A}lvarez-P{\'e}rez,~G. \latin{et~al.}  {Infrared permittivity of the biaxial van der Waals semiconductor $\alpha$-MoO$_3$ from near-and far-field correlative studies}. \emph{Advanced Materials} \textbf{2020}, \emph{32}, 1908176\relax
\mciteBstWouldAddEndPuncttrue
\mciteSetBstMidEndSepPunct{\mcitedefaultmidpunct}
{\mcitedefaultendpunct}{\mcitedefaultseppunct}\relax
\EndOfBibitem
\bibitem[Zhang \latin{et~al.}(2022)Zhang, Li, Chen, Ruta, Shao, Sternbach, McLeod, Sun, Xiong, Moore, Xu, Wu, Shabani, Zhou, Wang, Mooshammer, Ray, Wilson, Schuck, Dean, Pasupathy, Lipson, Xu, Zhu, Millis, Liu, Hone, and Basov]{Zhang2022nano}
Zhang,~S. \latin{et~al.}  Nano-spectroscopy of excitons in atomically thin transition metal dichalcogenides. \emph{Nature Comm.} \textbf{2022}, \emph{13}, 1--8\relax
\mciteBstWouldAddEndPuncttrue
\mciteSetBstMidEndSepPunct{\mcitedefaultmidpunct}
{\mcitedefaultendpunct}{\mcitedefaultseppunct}\relax
\EndOfBibitem
\bibitem[Keilmann and Hillenbrand(2004)Keilmann, and Hillenbrand]{Keilmann2004near}
Keilmann,~F.; Hillenbrand,~R. Near-field microscopy by elastic light scattering from a tip. \emph{Phil. Trans. R. Soc. A} \textbf{2004}, \emph{362}, 787--805\relax
\mciteBstWouldAddEndPuncttrue
\mciteSetBstMidEndSepPunct{\mcitedefaultmidpunct}
{\mcitedefaultendpunct}{\mcitedefaultseppunct}\relax
\EndOfBibitem
\bibitem[Fei \latin{et~al.}(2012)Fei, Rodin, Andreev, Bao, McLeod, Wagner, Zhang, Zhao, Thiemens, Dominguez, Fogler, Neto, Lau, Keilmann, and Basov]{Fei2012gate}
Fei,~Z.; Rodin,~A.; Andreev,~G.~O.; Bao,~W.; McLeod,~A.; Wagner,~M.; Zhang,~L.; Zhao,~Z.; Thiemens,~M.; Dominguez,~G.; Fogler,~M.~M.; Neto,~A. H.~C.; Lau,~C.~N.; Keilmann,~F.; Basov,~D.~N. Gate-tuning of graphene plasmons revealed by infrared nano-imaging. \emph{Nature} \textbf{2012}, \emph{487}, 82--85\relax
\mciteBstWouldAddEndPuncttrue
\mciteSetBstMidEndSepPunct{\mcitedefaultmidpunct}
{\mcitedefaultendpunct}{\mcitedefaultseppunct}\relax
\EndOfBibitem
\bibitem[Dai \latin{et~al.}(2014)Dai, Fei, Ma, Rodin, Wagner, McLeod, Liu, Gannett, Regan, Watanabe, Taniguchi, Thiemens, Dominguez, Neto, Zettl, Keilmann, Jarillo-Herrero, Fogler, and Basov]{Dai2014tunable}
Dai,~S. \latin{et~al.}  Tunable phonon polaritons in atomically thin van der Waals crystals of boron nitride. \emph{Science} \textbf{2014}, \emph{343}, 1125--1129\relax
\mciteBstWouldAddEndPuncttrue
\mciteSetBstMidEndSepPunct{\mcitedefaultmidpunct}
{\mcitedefaultendpunct}{\mcitedefaultseppunct}\relax
\EndOfBibitem
\bibitem[Yoxall \latin{et~al.}(2015)Yoxall, Schnell, Nikitin, Txoperena, Woessner, Lundeberg, Casanova, Hueso, Koppens, and Hillenbrand]{Yoxall2015direct}
Yoxall,~E.; Schnell,~M.; Nikitin,~A.~Y.; Txoperena,~O.; Woessner,~A.; Lundeberg,~M.~B.; Casanova,~F.; Hueso,~L.~E.; Koppens,~F.~H.; Hillenbrand,~R. Direct observation of ultraslow hyperbolic polariton propagation with negative phase velocity. \emph{Nature Photonics} \textbf{2015}, \emph{9}, 674--678\relax
\mciteBstWouldAddEndPuncttrue
\mciteSetBstMidEndSepPunct{\mcitedefaultmidpunct}
{\mcitedefaultendpunct}{\mcitedefaultseppunct}\relax
\EndOfBibitem
\bibitem[Ma \latin{et~al.}(2018)Ma, Alonso-Gonz{\'a}lez, Li, Nikitin, Yuan, Mart{\'\i}n-S{\'a}nchez, Taboada-Guti{\'e}rrez, Amenabar, Li, V{\'e}lez, Tollan, Dai, Zhang, Sriram, Kalantar-Zadeh, Lee, Hillenbrand, and Bao]{Ma2018plane}
Ma,~W. \latin{et~al.}  In-plane anisotropic and ultra-low-loss polaritons in a natural van der Waals crystal. \emph{Nature} \textbf{2018}, \emph{562}, 557--562\relax
\mciteBstWouldAddEndPuncttrue
\mciteSetBstMidEndSepPunct{\mcitedefaultmidpunct}
{\mcitedefaultendpunct}{\mcitedefaultseppunct}\relax
\EndOfBibitem
\bibitem[Taboada-Guti{\'e}rrez \latin{et~al.}(2020)Taboada-Guti{\'e}rrez, {\'A}lvarez-P{\'e}rez, Duan, Ma, Crowley, Prieto, Bylinkin, Autore, Volkova, Kimura, Kimura, Berger, Li, Bao, Gao, Errea, Hillenbrand, Martín-Sánchez, and Alonso-González]{Taboada2020broad}
Taboada-Guti{\'e}rrez,~J. \latin{et~al.}  Broad spectral tuning of ultra-low-loss polaritons in a van der Waals crystal by intercalation. \emph{Nat. Mater.} \textbf{2020}, \emph{19}, 964--968\relax
\mciteBstWouldAddEndPuncttrue
\mciteSetBstMidEndSepPunct{\mcitedefaultmidpunct}
{\mcitedefaultendpunct}{\mcitedefaultseppunct}\relax
\EndOfBibitem
\bibitem[Bylinkin \latin{et~al.}(2021)Bylinkin, Schnell, Autore, Calavalle, Li, Taboada-Guti{\`e}rrez, Liu, Edgar, Casanova, Hueso, Alonso-Gonzalez, Nikitin, and Hillenbrand]{Bylinkin2021real}
Bylinkin,~A.; Schnell,~M.; Autore,~M.; Calavalle,~F.; Li,~P.; Taboada-Guti{\`e}rrez,~J.; Liu,~S.; Edgar,~J.~H.; Casanova,~F.; Hueso,~L.~E.; Alonso-Gonzalez,~P.; Nikitin,~A.~Y.; Hillenbrand,~R. Real-space observation of vibrational strong coupling between propagating phonon polaritons and organic molecules. \emph{Nat. Photon.} \textbf{2021}, \emph{15}, 197--202\relax
\mciteBstWouldAddEndPuncttrue
\mciteSetBstMidEndSepPunct{\mcitedefaultmidpunct}
{\mcitedefaultendpunct}{\mcitedefaultseppunct}\relax
\EndOfBibitem
\bibitem[Woessner \latin{et~al.}(2015)Woessner, Lundeberg, Gao, Principi, Alonso-Gonz{\'a}lez, Carrega, Watanabe, Taniguchi, Vignale, Polini, Hone, Hillenbrand, and Koppens]{Woessner2015highly}
Woessner,~A.; Lundeberg,~M.~B.; Gao,~Y.; Principi,~A.; Alonso-Gonz{\'a}lez,~P.; Carrega,~M.; Watanabe,~K.; Taniguchi,~T.; Vignale,~G.; Polini,~M.; Hone,~J.; Hillenbrand,~R.; Koppens,~F. H.~L. Highly confined low-loss plasmons in graphene--boron nitride heterostructures. \emph{Nat. Mater.} \textbf{2015}, \emph{14}, 421--425\relax
\mciteBstWouldAddEndPuncttrue
\mciteSetBstMidEndSepPunct{\mcitedefaultmidpunct}
{\mcitedefaultendpunct}{\mcitedefaultseppunct}\relax
\EndOfBibitem
\bibitem[Walla \latin{et~al.}(2018)Walla, Wiecha, Mecklenbeck, Beldi, Keilmann, Thomson, and Roskos]{Walla2018anisotropic}
Walla,~F.; Wiecha,~M.~M.; Mecklenbeck,~N.; Beldi,~S.; Keilmann,~F.; Thomson,~M.~D.; Roskos,~H.~G. Anisotropic excitation of surface plasmon polaritons on a metal film by a scattering-type scanning near-field microscope with a non-rotationally-symmetric probe tip. \emph{Nanophotonics} \textbf{2018}, \emph{7}, 269--276\relax
\mciteBstWouldAddEndPuncttrue
\mciteSetBstMidEndSepPunct{\mcitedefaultmidpunct}
{\mcitedefaultendpunct}{\mcitedefaultseppunct}\relax
\EndOfBibitem
\bibitem[Kaltenecker \latin{et~al.}(2020)Kaltenecker, Krauss, Casses, Geisler, Hecht, Mortensen, Jepsen, and Stenger]{Kaltenecker2020mono}
Kaltenecker,~K.~J.; Krauss,~E.; Casses,~L.; Geisler,~M.; Hecht,~B.; Mortensen,~N.~A.; Jepsen,~P.~U.; Stenger,~N. Mono-crystalline gold platelets: a high-quality platform for surface plasmon polaritons. \emph{Nanophotonics} \textbf{2020}, \emph{9}, 509--522\relax
\mciteBstWouldAddEndPuncttrue
\mciteSetBstMidEndSepPunct{\mcitedefaultmidpunct}
{\mcitedefaultendpunct}{\mcitedefaultseppunct}\relax
\EndOfBibitem
\bibitem[Casses \latin{et~al.}(2022)Casses, Kaltenecker, Xiao, Wubs, and Stenger]{Casses2022quantitative}
Casses,~L.~N.; Kaltenecker,~K.~J.; Xiao,~S.; Wubs,~M.; Stenger,~N. Quantitative near-field characterization of surface plasmon polaritons on monocrystalline gold platelets. \emph{Opt. Express} \textbf{2022}, \emph{30}, 11181--11191\relax
\mciteBstWouldAddEndPuncttrue
\mciteSetBstMidEndSepPunct{\mcitedefaultmidpunct}
{\mcitedefaultendpunct}{\mcitedefaultseppunct}\relax
\EndOfBibitem
\bibitem[Basov \latin{et~al.}(2016)Basov, Fogler, and De~Abajo]{Basov2016polaritons}
Basov,~D.; Fogler,~M.; De~Abajo,~F.~G. {Polaritons in van der Waals materials}. \emph{Science} \textbf{2016}, \emph{354}, aag1992\relax
\mciteBstWouldAddEndPuncttrue
\mciteSetBstMidEndSepPunct{\mcitedefaultmidpunct}
{\mcitedefaultendpunct}{\mcitedefaultseppunct}\relax
\EndOfBibitem
\bibitem[Low \latin{et~al.}(2017)Low, Chaves, Caldwell, Kumar, Fang, Avouris, Heinz, Guinea, Martin-Moreno, and Koppens]{Low2017polaritons}
Low,~T.; Chaves,~A.; Caldwell,~J.~D.; Kumar,~A.; Fang,~N.~X.; Avouris,~P.; Heinz,~T.~F.; Guinea,~F.; Martin-Moreno,~L.; Koppens,~F. Polaritons in layered two-dimensional materials. \emph{Nat. Mater.} \textbf{2017}, \emph{16}, 182--194\relax
\mciteBstWouldAddEndPuncttrue
\mciteSetBstMidEndSepPunct{\mcitedefaultmidpunct}
{\mcitedefaultendpunct}{\mcitedefaultseppunct}\relax
\EndOfBibitem
\bibitem[Basov \latin{et~al.}(2021)Basov, Asenjo-Garcia, Schuck, Zhu, and Rubio]{Basov2021polariton}
Basov,~D.; Asenjo-Garcia,~A.; Schuck,~P.~J.; Zhu,~X.; Rubio,~A. Polariton panorama. \emph{Nanophotonics} \textbf{2021}, \emph{10}, 549--577\relax
\mciteBstWouldAddEndPuncttrue
\mciteSetBstMidEndSepPunct{\mcitedefaultmidpunct}
{\mcitedefaultendpunct}{\mcitedefaultseppunct}\relax
\EndOfBibitem
\bibitem[Mueller and Malic(2018)Mueller, and Malic]{Mueller2018exciton}
Mueller,~T.; Malic,~E. Exciton physics and device application of two-dimensional transition metal dichalcogenide semiconductors. \emph{npj 2D Materials and Applications} \textbf{2018}, \emph{2}, 1--12\relax
\mciteBstWouldAddEndPuncttrue
\mciteSetBstMidEndSepPunct{\mcitedefaultmidpunct}
{\mcitedefaultendpunct}{\mcitedefaultseppunct}\relax
\EndOfBibitem
\bibitem[Maier(2007)]{Maier2007plasmonics}
Maier,~S.~A. \emph{Plasmonics: Fundamentals and Applications}; Springer Science \& Business Media: New York, 2007\relax
\mciteBstWouldAddEndPuncttrue
\mciteSetBstMidEndSepPunct{\mcitedefaultmidpunct}
{\mcitedefaultendpunct}{\mcitedefaultseppunct}\relax
\EndOfBibitem
\bibitem[Xia \latin{et~al.}(2014)Xia, Wang, Xiao, Dubey, and Ramasubramaniam]{Xia2014two}
Xia,~F.; Wang,~H.; Xiao,~D.; Dubey,~M.; Ramasubramaniam,~A. Two-dimensional material nanophotonics. \emph{Nature Photonics} \textbf{2014}, \emph{8}, 899--907\relax
\mciteBstWouldAddEndPuncttrue
\mciteSetBstMidEndSepPunct{\mcitedefaultmidpunct}
{\mcitedefaultendpunct}{\mcitedefaultseppunct}\relax
\EndOfBibitem
\bibitem[Baranov \latin{et~al.}(2018)Baranov, Wersäll, Cuadra, Antosiewicz, and Shegai]{Baranov2018novel}
Baranov,~D.~G.; Wersäll,~M.; Cuadra,~J.; Antosiewicz,~T.~J.; Shegai,~T. Novel Nanostructures and Materials for Strong Light–Matter Interactions. \emph{ACS Photonics} \textbf{2018}, \emph{5}, 24--42\relax
\mciteBstWouldAddEndPuncttrue
\mciteSetBstMidEndSepPunct{\mcitedefaultmidpunct}
{\mcitedefaultendpunct}{\mcitedefaultseppunct}\relax
\EndOfBibitem
\bibitem[Byrnes \latin{et~al.}(2014)Byrnes, Kim, and Yamamoto]{Byrnes2014exciton}
Byrnes,~T.; Kim,~N.~Y.; Yamamoto,~Y. Exciton--polariton condensates. \emph{Nature Physics} \textbf{2014}, \emph{10}, 803--813\relax
\mciteBstWouldAddEndPuncttrue
\mciteSetBstMidEndSepPunct{\mcitedefaultmidpunct}
{\mcitedefaultendpunct}{\mcitedefaultseppunct}\relax
\EndOfBibitem
\bibitem[Sanvitto and K{\'e}na-Cohen(2016)Sanvitto, and K{\'e}na-Cohen]{Sanvitto2016road}
Sanvitto,~D.; K{\'e}na-Cohen,~S. The road towards polaritonic devices. \emph{Nature materials} \textbf{2016}, \emph{15}, 1061--1073\relax
\mciteBstWouldAddEndPuncttrue
\mciteSetBstMidEndSepPunct{\mcitedefaultmidpunct}
{\mcitedefaultendpunct}{\mcitedefaultseppunct}\relax
\EndOfBibitem
\bibitem[Denning \latin{et~al.}(2022)Denning, Wubs, Stenger, M\o{}rk, and Kristensen]{Denning2020cavity}
Denning,~E.~V.; Wubs,~M.; Stenger,~N.; M\o{}rk,~J.; Kristensen,~P.~T. Cavity-induced exciton localization and polariton blockade in two-dimensional semiconductors coupled to an electromagnetic resonator. \emph{Phys. Rev. Res.} \textbf{2022}, \emph{4}, L012020\relax
\mciteBstWouldAddEndPuncttrue
\mciteSetBstMidEndSepPunct{\mcitedefaultmidpunct}
{\mcitedefaultendpunct}{\mcitedefaultseppunct}\relax
\EndOfBibitem
\bibitem[B.~Iyer \latin{et~al.}(2022)B.~Iyer, Luan, Shinar, Shinar, and Fei]{Iyer2022nano}
B.~Iyer,~R.; Luan,~Y.; Shinar,~R.; Shinar,~J.; Fei,~Z. Nano-optical imaging of exciton–plasmon polaritons in WSe2/Au heterostructures. \emph{Nanoscale} \textbf{2022}, \emph{14}, 15663--15668\relax
\mciteBstWouldAddEndPuncttrue
\mciteSetBstMidEndSepPunct{\mcitedefaultmidpunct}
{\mcitedefaultendpunct}{\mcitedefaultseppunct}\relax
\EndOfBibitem
\bibitem[Jang \latin{et~al.}(2024)Jang, Menabde, Kiani, Heiden, Zenin, Mortensen, Tagliabue, and Jang]{Jang2024fourier}
Jang,~M.; Menabde,~S.~G.; Kiani,~F.; Heiden,~J.~T.; Zenin,~V.~A.; Mortensen,~N.~A.; Tagliabue,~G.; Jang,~M.~S. Fourier analysis of near-field patterns generated by propagating polaritons. \emph{arXiv preprint arXiv:2402.17225} \textbf{2024}, \relax
\mciteBstWouldAddEndPunctfalse
\mciteSetBstMidEndSepPunct{\mcitedefaultmidpunct}
{}{\mcitedefaultseppunct}\relax
\EndOfBibitem
\bibitem[Krauss \latin{et~al.}(2018)Krauss, Kullock, Wu, Geisler, Lundt, Kamp, and Hecht]{Krauss2018}
Krauss,~E.; Kullock,~R.; Wu,~X.; Geisler,~P.; Lundt,~N.; Kamp,~M.; Hecht,~B. Controlled Growth of High-Aspect-Ratio Single-Crystalline Gold Platelets. \emph{Cryst. Growth Des.} \textbf{2018}, \emph{18}, 1297--1302\relax
\mciteBstWouldAddEndPuncttrue
\mciteSetBstMidEndSepPunct{\mcitedefaultmidpunct}
{\mcitedefaultendpunct}{\mcitedefaultseppunct}\relax
\EndOfBibitem
\bibitem[Castellanos-Gomez \latin{et~al.}(2014)Castellanos-Gomez, Buscema, Molenaar, Singh, Janssen, Van Der~Zant, and Steele]{Castellanos2014deterministic}
Castellanos-Gomez,~A.; Buscema,~M.; Molenaar,~R.; Singh,~V.; Janssen,~L.; Van Der~Zant,~H.~S.; Steele,~G.~A. Deterministic transfer of two-dimensional materials by all-dry viscoelastic stamping. \emph{2D Materials} \textbf{2014}, \emph{1}, 011002\relax
\mciteBstWouldAddEndPuncttrue
\mciteSetBstMidEndSepPunct{\mcitedefaultmidpunct}
{\mcitedefaultendpunct}{\mcitedefaultseppunct}\relax
\EndOfBibitem
\bibitem[Bracewell(1986)]{Bracewell1986Fourier}
Bracewell,~R.~N. \emph{The Fourier Transform and its Applications - 3rd Ed.}; McGraw-Hill: New York, 1986\relax
\mciteBstWouldAddEndPuncttrue
\mciteSetBstMidEndSepPunct{\mcitedefaultmidpunct}
{\mcitedefaultendpunct}{\mcitedefaultseppunct}\relax
\EndOfBibitem
\bibitem[Maj{\'e}rus \latin{et~al.}(2018)Maj{\'e}rus, Dremetsika, Lobet, Henrard, and Kockaert]{Majerus2018electrodynamics}
Maj{\'e}rus,~B.; Dremetsika,~E.; Lobet,~M.; Henrard,~L.; Kockaert,~P. Electrodynamics of two-dimensional materials: Role of anisotropy. \emph{Physical Review B} \textbf{2018}, \emph{98}, 125419\relax
\mciteBstWouldAddEndPuncttrue
\mciteSetBstMidEndSepPunct{\mcitedefaultmidpunct}
{\mcitedefaultendpunct}{\mcitedefaultseppunct}\relax
\EndOfBibitem
\bibitem[McPeak \latin{et~al.}(2015)McPeak, Jayanti, Kress, Meyer, Iotti, Rossinelli, and Norris]{McPeak2015}
McPeak,~K.~M.; Jayanti,~S.~V.; Kress,~S. J.~P.; Meyer,~S.; Iotti,~S.; Rossinelli,~A.; Norris,~D.~J. Plasmonic Films Can Easily Be Better: Rules and Recipes. \emph{ACS Photonics} \textbf{2015}, \emph{2}, 326--333\relax
\mciteBstWouldAddEndPuncttrue
\mciteSetBstMidEndSepPunct{\mcitedefaultmidpunct}
{\mcitedefaultendpunct}{\mcitedefaultseppunct}\relax
\EndOfBibitem
\bibitem[Lebsir \latin{et~al.}(2022)Lebsir, Boroviks, Thomaschewski, Bozhevolnyi, and Zenin]{Lebsir2022ultimate}
Lebsir,~Y.; Boroviks,~S.; Thomaschewski,~M.; Bozhevolnyi,~S.~I.; Zenin,~V.~A. Ultimate Limit for Optical Losses in Gold, Revealed by Quantitative Near-Field Microscopy. \emph{Nano Letters} \textbf{2022}, \emph{22}, 5759--5764\relax
\mciteBstWouldAddEndPuncttrue
\mciteSetBstMidEndSepPunct{\mcitedefaultmidpunct}
{\mcitedefaultendpunct}{\mcitedefaultseppunct}\relax
\EndOfBibitem
\bibitem[Munkhbat \latin{et~al.}(2022)Munkhbat, Wróbel, Antosiewicz, and Shegai]{Munkhbat2022optical}
Munkhbat,~B.; Wróbel,~P.; Antosiewicz,~T.~J.; Shegai,~T.~O. Optical Constants of Several Multilayer Transition Metal Dichalcogenides Measured by Spectroscopic Ellipsometry in the 300–1700 nm Range: High Index, Anisotropy, and Hyperbolicity. \emph{ACS Photonics} \textbf{2022}, \emph{9}, 2398--2407\relax
\mciteBstWouldAddEndPuncttrue
\mciteSetBstMidEndSepPunct{\mcitedefaultmidpunct}
{\mcitedefaultendpunct}{\mcitedefaultseppunct}\relax
\EndOfBibitem
\bibitem[Zhan \latin{et~al.}(2013)Zhan, Shi, Dai, Liu, and Zi]{Zhan2013transfer}
Zhan,~T.; Shi,~X.; Dai,~Y.; Liu,~X.; Zi,~J. Transfer matrix method for optics in graphene layers. \emph{Journal of Physics: Condensed Matter} \textbf{2013}, \emph{25}, 215301\relax
\mciteBstWouldAddEndPuncttrue
\mciteSetBstMidEndSepPunct{\mcitedefaultmidpunct}
{\mcitedefaultendpunct}{\mcitedefaultseppunct}\relax
\EndOfBibitem
\bibitem[Geisler \latin{et~al.}(2019)Geisler, Cui, Wang, Rindzevicius, Gammelgaard, Jessen, Gon{\c{c}}alves, Todisco, B{\o}ggild, Boisen, Wubs, Mortensen, Xiao, and Stenger]{Geisler2019single}
Geisler,~M.; Cui,~X.; Wang,~J.; Rindzevicius,~T.; Gammelgaard,~L.; Jessen,~B.~S.; Gon{\c{c}}alves,~P. A.~D.; Todisco,~F.; B{\o}ggild,~P.; Boisen,~A.; Wubs,~M.; Mortensen,~N.~A.; Xiao,~S.; Stenger,~N. {Single-crystalline gold nanodisks on WS$_2$ mono- and multilayers for strong coupling at room temperature}. \emph{ACS Photonics} \textbf{2019}, \emph{6}, 994--1001\relax
\mciteBstWouldAddEndPuncttrue
\mciteSetBstMidEndSepPunct{\mcitedefaultmidpunct}
{\mcitedefaultendpunct}{\mcitedefaultseppunct}\relax
\EndOfBibitem
\bibitem[Gon{\c{c}}alves \latin{et~al.}(2018)Gon{\c{c}}alves, Bertelsen, Xiao, and Mortensen]{Goncalves2018plasmon}
Gon{\c{c}}alves,~P.; Bertelsen,~L.; Xiao,~S.; Mortensen,~N.~A. Plasmon-exciton polaritons in two-dimensional semiconductor/metal interfaces. \emph{Physical Review B} \textbf{2018}, \emph{97}, 041402\relax
\mciteBstWouldAddEndPuncttrue
\mciteSetBstMidEndSepPunct{\mcitedefaultmidpunct}
{\mcitedefaultendpunct}{\mcitedefaultseppunct}\relax
\EndOfBibitem
\bibitem[Boroviks \latin{et~al.}(2022)Boroviks, Lin, Zenin, Ziegler, Dellith, Gon{\c{c}}alves, Wolff, Bozhevolnyi, Huang, and Mortensen]{Boroviks2022extremely}
Boroviks,~S.; Lin,~Z.-H.; Zenin,~V.~A.; Ziegler,~M.; Dellith,~A.; Gon{\c{c}}alves,~P.; Wolff,~C.; Bozhevolnyi,~S.~I.; Huang,~J.-S.; Mortensen,~N.~A. Extremely confined gap plasmon modes: when nonlocality matters. \emph{Nature Communications} \textbf{2022}, \emph{13}, 1--8\relax
\mciteBstWouldAddEndPuncttrue
\mciteSetBstMidEndSepPunct{\mcitedefaultmidpunct}
{\mcitedefaultendpunct}{\mcitedefaultseppunct}\relax
\EndOfBibitem
\end{mcitethebibliography}

\end{document}